\documentclass[journal]{IEEEtran}
\usepackage{amsmath}
\usepackage{amssymb}
\usepackage{graphicx}
\usepackage{subfig}
\usepackage{float}
\usepackage{cite}
\usepackage{diagbox}
\usepackage{tabularx,booktabs}
\usepackage{cellspace}
\usepackage{color}
\usepackage{caption}
\usepackage{varwidth}
\usepackage{afterpage}
\usepackage{optidef}
\DeclareCaptionFormat{myformat}{%
  \begin{varwidth}{\linewidth}%
    \centering
    #1#2#3%
  \end{varwidth}%
}

\DeclareMathOperator*{\minimize}{minimize}
\DeclareMathOperator*{\subjectto}{s.t.}

\usepackage{amsthm}
\newtheorem{theorem}{Theorem}

\newtheorem{definition}{Definition}
\usepackage[ruled,vlined]{algorithm2e}
\usepackage{multirow}
\newcolumntype{C}{>{\centering\arraybackslash}X} 
\graphicspath{ {\string./Figures/} }
\ifCLASSINFOpdf
\else
\fi

\usepackage{xcolor}

\newif\ifdraft \drafttrue

\begin{document}
\title{Active Privacy-Utility Trade-off Against \\ Inference in Time-Series Data Sharing}

\author{Ecenaz Erdemir,~\IEEEmembership{Student Member,~IEEE,}
        Pier Luigi Dragotti,~\IEEEmembership{Fellow,~IEEE,}
        \\ and~Deniz G\"{u}nd\"{u}z,~\IEEEmembership{Fellow,~IEEE}
\sloppy
\thanks{The authors are with the Department
of Electrical and Electronic Engineering, Imperial College London, London SW7 2AZ, U.K., (e-mail: \{e.erdemir17, p.dragotti, d.gunduz\}@imperial.ac.uk).}
}

\maketitle

\begin{abstract}
\label{sec:abstract}
Internet of things (IoT) devices, such as smart meters, smart speakers and activity monitors, have become highly popular thanks to the services they offer. However, in addition to their many benefits, they raise privacy concerns since they share fine-grained time-series user data with untrusted third parties.
In this work, we consider a user releasing her data containing personal information in return of a service from an honest-but-curious service provider (SP). We model user’s personal information as two correlated random variables (r.v.'s), one of them, called the \textit{secret variable}, is to be kept private, while the other, called the \textit{useful variable}, is to be disclosed for utility. We consider active sequential data release, where at each time step the user chooses from among a finite set of release mechanisms, each revealing some information about the user's personal information, i.e., the true values of the r.v.'s, albeit with different statistics. The user manages data release in an online fashion such that the maximum amount of information is revealed about the latent useful variable as quickly as possible, while the confidence for the sensitive variable is kept below a predefined level. For privacy measure, we consider both the probability of correctly detecting the true value of the secret and the mutual information (MI) between the secret and the released data.
We formulate both problems as partially observable Markov decision processes (POMDPs), and numerically solve them by advantage actor-critic (A2C) deep reinforcement learning (DRL). We evaluate the privacy-utility trade-off (PUT) of the proposed policies on both the synthetic data and \textit{smoking activity dataset}, and show their validity by testing the activity detection accuracy of the SP modeled by a long short-term memory (LSTM) neural network.
\end{abstract}

\vspace{-0.1cm}
\begin{IEEEkeywords}
Inference privacy, time-series privacy, privacy funnel, active learning, actor-critic deep reinforcement learning, human activity recognition.
\end{IEEEkeywords}

\IEEEpeerreviewmaketitle

\vspace{-0.55cm}
\section{Introduction}

\makeatletter{\renewcommand*{\@makefnmark}{}\footnotetext{This work was partially supported by the the European Research Council through project BEACON (grant number 677854) and UK EPSRC through CHIST-ERA project CONNECT (CHISTERA-18-SDCDN-001, EPSRC-EP/T023600/1).}\makeatother}
\vspace{-0.1cm}
\IEEEPARstart{R}{ecent} advances in Internet of things (IoT) devices and services have increased their usage in a wide range of areas, such as health and activity monitoring, location-based services, smart speakers and smart metering.
Moreover, most service providers encourage the users to share their personal data in return for better user experience. For instance, the users can benefit from personalized dietary tips as a result of sharing their Fitbit activity, while they can receive hotel, restaurant or bar recommendations if they share their location.
However, in most of these applications, data collected by IoT devices contain sensitive personal information about the users. 
The concerning fact is that as soon as the user's raw data is sent to the service provider's cloud, the sensitive information can be inferred, misused or leaked through security vulnerabilities
even if the service provider and/or the communication link
are trusted third parties.
For example, chronic illnesses, disabilities, daily habits and psychological state can be revealed by health monitoring systems \cite{ECG,Health}, while presence at home and states of home appliances can be inferred from collected smart meter (SM) data \cite{GiulioSPmag}.
Hence, privacy is an important concern for the adoption of many IoT services, and there is a growing demand from consumers to keep their personal information private against malicious attackers and/or untrusted service providers (SPs), while preserving the utility obtained from these IoT services. Privacy has been widely studied in the literature \cite{DiffPrivacyMarkov,erdemir2022privacyaware,statInf,SkoglundEpsPriv,BizBook,WIFS,TIFS,ICASSP,FunnelLimits,DenizTobias,KLprivacy,APUT_HTAdv}, and a vast number of privacy measures have been introduced, including differential privacy \cite{DiffPrivacyMarkov}, mutual information (MI) \cite{statInf, SkoglundEpsPriv,BizBook,ICASSP,WIFS,TIFS,FunnelLimits}, total variation distance \cite{Borzoo}, maximal leakage \cite{SankarAlphaMaxLeak,MaxLeak}, and guessing leakage \cite{HamedBorzoo}, to count a few.
In this paper, we study the privacy-utility trade-off (PUT) of time-series data sharing in the presence of a third party which tries to infer the user's sensitive information from the released data.

\vspace{-0.5cm}
\subsection{Related Work}
\vspace{-0.1cm}
Privacy for time-series data sharing and its applications to various domains have been extensively studied \cite{Timeseries,Health_kanon2,Health_DifP,Shokri_kanony,Shokri_single,InfoTheo_single,SM_DifP1,SM_DifP2,GiulioRED_ESD,Ravi,Trace1,WIFS, TIFS,Giulio,ICASSP,BizBook,APUT_HTAdv}. Most of these works focus on protecting the privacy of a single data point, e.g., the current measurement \cite{Shokri_kanony,Shokri_single,InfoTheo_single,SM_DifP1,SM_DifP2,GiulioRED_ESD}. However, causal relations in time-series data require taking into account more than single data point privacy. Individual measurements taken at each time instance, such as electrocardiogram (ECG), body temperature, physical activity, location, weather forecast, account balance and SM readings, are temporally correlated and the strategies focusing on the privacy of a single data point might reveal sensitive information about the past or future measurements.

Among those that consider temporal correlations, most existing works focus on the privacy of the time-series measurements rather than hiding latent sensitive attributes \cite{WIFS,TIFS,Ravi,Trace1,GiulioRED_ESD,SM_DifP1,SM_DifP2}. In the location sharing scenario, sensitive information is the time-series data itself and the utility loss can be measured by data distortion, whereas in many applications, the user might be interested in hiding an underlying sensitive hypothesis. For instance, the user's presence at home or favorite TV channel can be inferred from SM readings, while her sensitive daily habits can be revealed to the SP through the sensors of a wearable device. 
Inference privacy protects user's data from an adversary’s attempt to deduce sensitive information from an underlying distribution \cite{Funnel,FunnelLimits,HamedBorzoo,Inf_Data_P,PPAN,Erdogdu_TimeSeriesISIT,puff3,APUT_HTAdv}. These techniques perform well against inference attacks, in which the adversary aims at detecting the user's underlying private information with high confidence \cite{Trace1}.
PUT between two correlated sensitive and useful r.v.'s has also been studied under \textit{privacy funnel} \cite{Funnel}, which is closely related to \textit{information bottleneck} introduced in \cite{Tishby}. In privacy funnel approaches \cite{Funnel,FunnelLimits,HamedBorzoo,Inf_Data_P,PPAN,Erdogdu_TimeSeriesISIT,puff3}, the goal is to conceal the sensitive information from SP's inference while gaining enough utility from the useful information, where both the utility and the privacy leakage are measured by MI. 
However, \cite{Funnel,FunnelLimits,Inf_Data_P,PPAN} consider independent data without temporal correlations, hence, these approaches are not suitable for temporally correlated time-series data.


Differential privacy (DP), k-anonymity, information theoretic metrics and the SP's error probability are commonly used as privacy measures \cite{Timeseries,Health_kanon2,Health_DifP,Shokri_kanony,Shokri_single,InfoTheo_single,SM_DifP1,SM_DifP2,GiulioRED_ESD,Ravi,Trace1,WIFS,ICASSP,Giulio,BizBook, ShokriQuantify, APUT_HTAdv}.
By definition, DP prevents the SP from inferring the current data of the user, even if the SP has the knowledge of all the remaining data points. K-anonymity ensures that sensitive data is indistinguishable from at least $k-1$ other data points. However, DP and k-anonymity are meant to ensure the privacy of a single point in a time-series and do not consider temporal correlations. 
As an intermediate framework between complete independence and complete correlation, \textit{pufferfish privacy} considers low temporal correlations in time-series \cite{puff3}. However, the mechanism focuses on the privacy around the possible current data, which might ignore the inference from future and past values. 

In the literature, several papers on DP consider temporal correlations \cite{DiffPrivacyMarkov}. However, these usually follow myopic correlations with the current data due to the utility loss concerns raising from high noise. For example, in \cite{Health_DifP}, physiological measurements are obfuscated before reporting to an SP for PUT. Instead of the entire time-series history, a selected temporal section of the sensor data is considered, and solved by using dynamic programming (DyP) and greedy algorithm. 
In \cite{SM_DifP2}, DP in a SM with a rechargeable battery is achieved by adding noise to the meter readings before reporting to an SP. In order to guarantee DP, the perturbation must be independent of the battery state of charge. However, for a finite capacity battery, the energy management system cannot provide the amount of noise required for preserving privacy.

Information-theoretic privacy (ITP) considers the statistics of the entire time-series in terms of temporal correlations, and study privacy mechanisms that allow arbitrary stochastic transformations of data samples. However, DP considers only a single data point privacy. This disadvantage of DP is partially eliminated by group DP and pufferfish privacy, which take some degree of temporal correlations into account by considering DP of multiple neighboring data points and adding a fixed type of i.i.d. random noise to the samples. However, these approaches lead to significant utility loss, since each specific privacy mechanism in DP limits the type of noise added, e.g., Gaussian, Laplacian, and etc. On the other hand, ITP has no such restriction in the stochastic transformations that are applied to the data samples.
This is one of the biggest advantage of ITP over DP and pufferfish privacy where only some degree of temporal correlations are taken into account, and a fixed type of i.i.d. random noise is added for privacy. 

In \cite{AshishInfoTheoP}, an SM system is considered assuming Markovian energy demands. Privacy is achieved by filtering the energy demand with the help of a rechargeable battery. ITP problem is formulated as an MDP, and the minimum leakage is obtained numerically through DyP, while a single-letter expression is obtained for an i.i.d. demand. This approach is extended to the scenario with a renewable energy source in \cite{Giulio}. In \cite{ParvMDP}, PUT is examined with a rechargeable battery. Due to Markovian demand and price processes, the problem is formulated as a partially observable MDP with belief-dependent rewards ($\rho$-POMDP), and solved by DyP for infinite-horizon. In \cite{ICASSP}, PUT is characterized numerically by DyP for a special energy generation process. 

In \cite{Erdogdu_TimeSeriesISIT}, PUT of time-series data is considered in both online and offline settings. A user continuously releases data samples which are correlated with its private information, and in return obtains utility from an SP. The proposed schemes are cast as convex optimization problems and solved under hidden Markov model assumption. The simulation results are provided for binary time-series data for a finite time horizon. However, the dimensions of the optimization problems in both schemes grow exponentially with time and the number of sample states. Therefore, in a setting when fine-grained sensor data is considered for a long time horizon, computational complexity of the proposed schemes is very high.

ITP and utility for location sharing is studied in \cite{WIFS}, and extended to generic time-series data release in \cite{TIFS}. The user follows a history-dependent online data release policy by minimizing the MI between the real and modified location trajectories subject to a distortion constraint. Effectiveness of the proposed approach against myopic policies and its application to GeoLife GPS trajectory dataset are presented through numerical simulations.

Privacy metrics based on the SP's error probability focus on concealing the true realization of the sensitive information. In \cite{DenizTobias}, the goal is to increase the fidelity of the shared data quantified through an additive distortion measure, while guaranteeing privacy in an online manner. Privacy leakage is measured by the error probability of the SP in detecting the distribution of the underlying data samples. In \cite{APUT_HTAdv}, the user shares her time-series data, which intrinsically contains correlated sensitive and useful information, with an untrusted SP in an online fashion. The goal is to maximize the confidence in the true useful variable for utility, while keeping the confidence in the sensitive r.v. below a pre-defined level. 


\vspace{-0.4cm}
\subsection{Contributions}
In this paper, we consider an active learning scenario for PUT against an honest-but-curious SP. We assume that a user wants to share the ``useful'' part of her data with the SP. However, the SP might also deduce user's ``secret" information from the shared time-series data (e.g., location, heartbeat, temperature or energy consumption). We model the user's secret and useful data as correlated discrete r.v.'s. The user's goal is to prevent the secret from being accurately detected by the SP while revealing the useful data accurately for utility. 

Differently from the existing works \cite{statInf,SkoglundEpsPriv,Borzoo,SankarAlphaMaxLeak,HamedBorzoo,Funnel,PPAN}, which typically consider a time-independent data release problem, we consider a discrete time system, and assume that the user can actively choose from among a finite number of data release mechanisms (DRMs) at each time. While each measurement reveals some information about user's latent states, we assume that each DRM has different measurement characteristics, i.e., conditional probability distributions. User's objective is to choose a DRM at each time in an online fashion to reveal the value of the useful r.v. as quickly as possible to maximize her utility while keeping the leakage of the sensitive information below a prescribed value. As privacy measure, we consider both the SP's confidence in the secret and the MI between the secret r.v. and the observations. 

Our problem is similar to time-series data privacy in the literature \cite{Giulio,BizBook,ICASSP,WIFS,TIFS,Ravi, GiulioRED_ESD}, where the objective is to minimize privacy leakage by modifying the original time-series data while constraining the utility loss. However, in this work, the user selects from among multiple DRMs in an online fashion rather than modifying the non-causally available time-series data.
Similar time-series data release problems are also considered in \cite{DenizTobias},\cite{APUT_HTAdv} and \cite{SankarHTP}. However, \cite{DenizTobias} considers the PUT of a binary secret r.v. in an asymptotic regime, while \cite{SankarHTP} considers binary as well as M-ary r.v.'s for an offline scenario using semi-definite programming, which has high computational complexity when fine-grained data is considered. Data release history is taken into account for  M-ary r.v.'s in \cite{APUT_HTAdv}; however, time aspect is not considered in the PUT objective.

We introduce a sequential data release policy that minimizes the SP's error probability on the useful r.v. subject to a constraint on the SP's confidence in the true secret. Besides confidence-based privacy, we also consider MI-based privacy which keeps the total MI between the secret r.v. and the shared data below a certain level. Note that MI-based privacy does not necessarily prevent the detection of the true secret value; instead, it limits the information leakage in an average sense.

We consider data release policies which take the entire release
history into account, and recast the problem under both privacy measures as a POMDP. POMDPs can be represented as continuous state belief-MDPs; however, finding optimal policies for continuous state and action spaces is a PSPACE-hard problem \cite{PSPACEhard}. Solving these MDPs with fine discretization causes an increase in the state space; hence, the complexity. Therefore, after identifying the structure of the optimal policy, we use a method called advantage actor-critic (A2C) deep reinforcement learning (DRL) to evaluate our continuous state and action probability space MDP numerically. We also use variational representations for MI estimation through neural networks.
Finally, we examine the performances of the proposed policies in human activity privacy scenario, in which we use both synthetic data and smartwatch sensor readings from \textit{smoking activity dataset} \cite{smokingDataset}. 
We compare the privacy levels achieved by the proposed policies using an SP that predicts the true values for useful data and secret from the shared observation history. The SP is represented by a long short-term
memory (LSTM) neural network.

Our contributions are summarized as follows:
\begin{itemize}
    \item We propose an active learning framework for PUT in online sharing of time-series data.
    \item We propose a data sharing policy for optimal PUT against an SP performing sequential Bayesian inference.
    \item We propose another policy based on privacy measured by MI between the sensitive information and the released data history against average-case adversaries.
    \item We recast the active time-series data release problem for PUT as a POMDP, and evaluate both policies numerically using A2C-DRL for human activity privacy application.
\end{itemize}

The remainder of the paper is organized as follows. We present the problem formulation and POMDP approach in Sections \ref{sec:ProblemFormulation} and \ref{sec:POMDPFormulation}, respectively. MI-based privacy is introduced in Section \ref{sec:MI}, and data-driven evaluation for human activity privacy is presented in Section \ref{sec:NumericalResults}. Finally, we conclude our work in Section \ref{sec:Conclusion}.

\vspace{-0.4cm}
\section{Problem Statement}
\label{sec:ProblemFormulation}
We consider a user that wants to share her data with an honest-but-curious SP in return of utility. The data reveals information about two underlying latent variables; one represents the user's sensitive information, called the \textit{secret}, while the other is non-sensitive useful part, and is intentionally disclosed for utility. The user's goal is to release her data such that the SP can quickly detect the non-sensitive information with minimum error, while keeping his confidence in the secret r.v. below a predefined level. 


Fig. \ref{fig:SystemModel} shows an illustration of the system model with three DRMs. 
Let $\mathcal{S}=\{0,1, \dots, N-1\}$ and $\mathcal{U}=\{0,1, \dots, M-1\}$ be the finite sets of the hypotheses represented by the r.v.'s $S \in \mathcal{S}$ for the secret and $U \in \mathcal{U}$ for the non-sensitive useful information, respectively. Consider a finite set $\mathcal{A}$ of different DRMs available to the user, each modeled with a different statistical relation with the underlying hypotheses. For example, in the case of a user sharing activity data, e.g., Fitbit records, set $\mathcal{A}$ may correspond to different types of sensor measurements the user may share. Useful information the user wants to share may be the exercise type, while the sensitive information can be various daily habits. Similarly, in the case of smart meter readings, the useful information might be ON/OFF state of home appliances for smart power scheduling whereas the sensitive information might be the types of TV channels the user watches. We assume that the data revealed at time $t$, $Z_t$, is generated by an independent realization of a conditional probability distribution that depends on the true hypotheses and the chosen DRM $A_t \in \mathcal{A}$, denoted by $q(Z_t|A_t,S,U)$.

{\begin{figure}[pt]
\centering
\includegraphics[width=8.8cm]{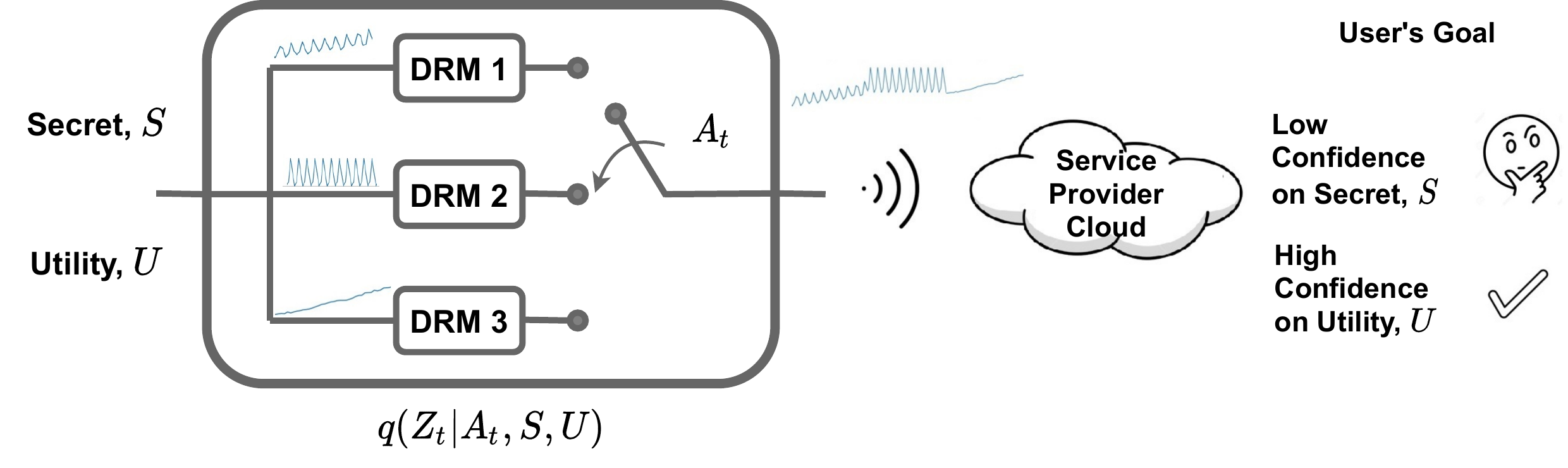}
\caption{System model for active PUT against the SP. } 
\label{fig:SystemModel}
\end{figure}}

The user's goal is to disclose $U$ through the released data $Z_t$, while keeping the SP's confidence in $S$ below a certain threshold. Let $\tau$ be the time that the SP is confident enough about the true useful variable and makes a declaration. This is also the time at which the user stops releasing data, since $U$ is already disclosed to the SP at the desired confidence level. The objective of the problem is to find a sequence of actions $\{A_0, \dots, A_{\tau-1}\}$, a stochastic stopping time $\tau$ and a declaration rule $d:\mathcal{A}^{\tau-1} \times \mathcal{Z}^{\tau-1} \rightarrow \mathcal{U}$ that collectively solve the following optimization problem:
\begin{mini}|l|
	  {A_0, \dots, A_{\tau-1}, d}{\mathbb{E}[\tau]+\lambda P_{err}(u) }{}{}
	  \addConstraint{\mathcal{C}_t(s)}{< L_{B}, }{ \forall t \leq \tau, \forall s \in \mathcal{S}}
	  \label{eqn:ProblemP}
\end{mini}
where $P_{err}(u)=P(d(A^{\tau-1},Z^{\tau-1})\neq u)$ is the error probability of making wrong declaration for the true value $u \in \mathcal{U}$; $\mathcal{C}_t(s)$ is the SP's instantaneous confidence in the true sensitive value $s \in \mathcal{S}$, which is a probability distribution on $s$ given the observation history, i.e., $P(S=s|A^{\tau-1},Z^{\tau-1})$; $L_{B}$ is a scalar of user's choice; and the expectation is taken over the action and observation distributions as well as the initial distributions of the r.v.'s. Here, by adjusting $\lambda$, we can trade-off between the speed of declaration and the SP's accuracy.

For our theoretical results, we assume that the observation statistics $q(Z_t|A_t,S,U), \forall A_t \in \mathcal{A}$, and the employed DRM $A_t$ are known by both the user and the SP. Later, we will also consider real datasets with unknown data distributions in our simulations. To maximally confuse the SP, the user selects action $A_t$ with a probability distribution $\pi(A_t|Z^{t-1}, A^{t-1})$ conditioned on the SP's observation history up to that time, $\{Z^{t-1}, A^{t-1}\}$. In this work, we assume that the true values $s$ and $u$ are unknown to all the parties involved.

\vspace{-0.3cm}
\section{POMDP Formulation}
\label{sec:POMDPFormulation}
The above PUT can be recast as a POMDP with partially observable static states $\{S,U\} \in \mathcal{S}\times\mathcal{U}$, actions $A_t\in \mathcal{A} \cup \{d\}$, and noisy observations $Z_t\in \mathcal{Z}$. POMDPs can be reformulated as belief-MDP with compact yet uncountable belief state, and solved using classical MDP solution methods.
We will follow this approach, and introduce SP's belief to determine the state variable in three steps. Firstly, we define the belief of the SP on $S$ and $U$ after he observes $\{Z^{t-1},A^{t-1}\}$ by
\begin{align}
    \hspace{-0.1cm}
    \beta_t(s,u)=P(S\hspace{-0.1cm}=s,U\hspace{-0.1cm}=u|Z^{t-1}=z^{t-1},A^{t-1}=a^{t-1})
\end{align}
over the belief space $\mathbb{P}(\mathcal{B}):=\{\beta_t \in [0,1]^{M \times N}:\sum_{s \in \mathcal{S}, u \in \mathcal{U}}\beta_t(s,u)=1\}$, where the marginal beliefs are represented by $\beta_t(u):=\sum_{s \in \mathcal{S}}\beta_t(s,u)$ and $\beta_t(s):=\sum_{u \in \mathcal{U}}\beta_t(s,u)$, respectively. The SP's confidence that $S=s$ at time $t$ is represented by $\mathcal{C}_t(s):=\beta_t(s)$.
The user's action probabilities become conditioned on the belief distribution, i.e., $\pi(A_t=a_t|\beta_t)$, while the observation probabilities are the same as before.
Secondly, we introduce a new state $F_{B}:=\{\max\limits_{s \in \mathcal{S}}\beta_t(s) \geq L_{B}: \beta_t \in \mathbb{P}(\mathcal{B})\}$ for $F_{B}\subseteq\mathbb{P}(\mathcal{B})$, called the \textit{forbidden-state}, which represents the condition where the constraint in (\ref{eqn:ProblemP}) is violated. With slight abuse of notation, we will use $F_B$ to denote both the forbidden state of the system and the set of belief states that fall into this state. $F_{B}$ is ideally an infinite cost state; however, in practice, we assume it has a large-cost. As the third step of defining the state space, we include a terminal state to fully characterize the state in which the user stops sharing her data with the SP.
We assume that after the user makes the stopping decision, the system goes to a terminal state, denoted by $F_T$, and remains there forever. This makes the problem an episodic MDP. Consequently, the state space becomes $\mathcal{X}=\mathbb{P}(\mathcal{B}) \cup \{F_T\}$. 

We always refer the time independent expression of belief, i.e., $\beta$, as the current belief state. The optimal expected total cost of our problem is defined as follows:
\begin{definition}
For all $\beta \in \mathbb{P}(\mathcal{B})$, let the optimal value function $V^*(\beta)$ represent the optimal expected cost of problem (\ref{eqn:ProblemP}), given the initial belief $\beta$. That is,
\begin{equation}
    V^*(\beta):=\min\{\mathbb{E}[\tau]+\lambda P_{err}(u)\},
\end{equation}
where the minimization is with respect to $\tau$, DRMs, and the declaration rule $d$. \label{def:OptVal}
\end{definition}

Optimal expected total cost for active PUT against an SP can be obtained by evaluating $V^*$ at the initial belief. This can be done by solving a DyP problem. After a single observation $\{z_t,a_t\}$, the SP updates its belief by Bayes' rule as follows:
\begin{align}
    \Phi(\beta_t,z_t,a_t) = \frac{q(z_t|a_t,s,u)\beta_t(s,u)}{\sum\limits_{\tilde{s},\tilde{u}}q(z_t|a_t,\tilde{s},\tilde{u})\beta_t(\tilde{s},\tilde{u})},
    \label{BeliefUpdate}
\end{align}
where the function $\Phi(\beta_t,z_t,a_t)$ represents the next belief state $\beta_{t+1}(s,u)$ in terms of the current belief, the action and the observation.
We define a Markov operator $\mathbb{T}^a$ for action $a$, such that for any measurable function $V:\mathbb{P}(\mathcal{B}) \rightarrow \mathbb{R}$,
\begin{equation}
    (\mathbb{T}^a V)(\beta):=\int V(\Phi(\beta,z,a))\sum_{s,u}q(z|a,s,u)\beta(s,u)dz.
\end{equation}
For any state $\beta \in \mathbb{P}(\mathcal{B})$, the user's data release action $a \in \mathcal{A}$ under the optimal policy results in an expected total cost of $1+(\mathbb{T}^a V^*)(\beta)$, where time spent by the user for data release is represented by cost 1, and $(\mathbb{T}^a V^*)(\beta)$ is the expected future value of $V^*$. On the other hand, the user's stopping decision $d$ results in error probability of the declaration of true useful value $u$ with penalty $\lambda$, i.e., $\lambda P_{err}(u):=\lambda (1-\beta(u))$. Solution for the optimal $V^*$ is formalized by the following theorem.
\begin{theorem}\cite{Bertsekas2}
The optimal $V^*$ for $\beta \in \mathbb{P}(\mathcal{B})$ satisfies the fixed point equation:
\begin{equation}
   \hspace{-0.02cm} V^*(\beta)=\min\{1+\min\limits_{a \in \mathcal{A}}(\mathbb{T}^a V^*)(\beta), \min\limits_{u \in \mathcal{U}}\lambda(1-\beta(u))\}. \label{eqn:optimalpolicy}
\end{equation}
\label{prop:fixpoint}
\end{theorem}

\vspace{-0.4cm}
\begin{definition}
\label{def:MarkovPolicy}
Let a Markov stationary policy $\pi$ be a stochastic kernel from the state space to the action space, including the stopping action, which determines the stopping time $\tau$, i.e., $\boldsymbol{\Pi} := \mathbb{P}(\mathcal{B}) \rightarrow \mathcal{A} \cup \{d\}$. That is, the probability of choosing DRM $a$ under policy $\pi$ at state $\beta$ is denoted by $\pi(a|\beta)$.
\end{definition}

Following from Corollary 9.12.1 in \cite{Bertsekas2}, DyP equation (\ref{eqn:optimalpolicy}) characterizes the optimal deterministic stationary policy $\pi^*$ for $\beta \in \mathbb{P}(\mathcal{B})$. The intuition behind Theorem \ref{prop:fixpoint} is that the user's data release action $a^*=\text{arg}\min_{a \in \mathcal{A}}T^a(V^*)(\beta)$ is the least costly action with cost $1+\min_{a \in \mathcal{A}}T^a(V^*)(\beta)$, unless choosing the stopping action $d$ and letting the SP make a decision for $u$ is less costly, i.e., $\lambda(1-\beta(u))$. We also ensure that for any two hypotheses $u, u'\in\mathcal{U}$, $u \neq u'$, there exists an action $a \in \mathcal{A}$, such that $D(q(z|a,s,u)||q(z|a,s,u')) > 0, \forall s \in \mathcal{S}$, where $\text{D}(\cdot||\cdot)$ denotes the Kullback-Leibler (KL) divergence. That is, hypotheses $u$ and $u'$ are distinguishable all the time, such that (\ref{eqn:ProblemP}) has a meaningful solution.

\begin{theorem}
 Suppose there exists a parameter $C_T > 0$, e.g., time cost, and a functional $V:\mathbb{P}(\mathcal{B}) \rightarrow \mathbb{R}_+$ such that for all belief states $\beta \in \mathbb{P}(\mathcal{B})$,
\begin{equation}
  \hspace{-0.03cm}  V(\beta) \leq \min\{C_T+\min\limits_{a \in \mathcal{A}}(\mathbb{T}^a V^*)(\beta), \min\limits_{u \in \mathcal{U}}\lambda C_T(1-\beta(u))\}. \label{eqn:thm1}
\end{equation}
Then $V^*(\beta) \geq \frac{1}{C_T}V(\beta)$ for all $\beta \in \mathbb{P}(\mathcal{B})$.
\label{thm:convergence}
\end{theorem}
See Appendix \ref{apx:ThmProof} for the proof of Theorem \ref{thm:convergence}.

Theorem \ref{thm:convergence} provides a lower bound for a fixed-point expression of $V^*$. However, it is difficult to calculate the real value of $V^*$ and solve the DyP equation over a continuous belief space. Hence, we solve (\ref{eqn:ProblemP}) numerically using an RL approach to obtain a good approximation.
Due to the belief-based privacy constraint, we call our policy \textit{belief-privacy data release policy} (belief-PDRP), $\pi_{B}$.
We define an instantaneous cost function for current state $x$ and action $a \in \mathcal{A} \cup \{d\}$ as
\begin{equation}
 \hspace{-0.1cm}   c^{\pi_{B}}(x,a) \hspace{-0.1cm} = \hspace{-0.1cm}
    \begin{cases}
      1, \hspace{-0.3cm}  & \hspace{-0.2cm} \text{if}\ x=\beta \in \mathbb{P}(\mathcal{B})\hspace{-0.1cm} \setminus \hspace{-0.1cm} F_{B}, a \in \mathcal{A} \\
      \min \limits_{u \in \mathcal{U}}(1-\beta(u))\lambda, & \hspace{-0.2cm} \text{if}\ x=\beta \in \mathbb{P}(\mathcal{B})\hspace{-0.1cm} \setminus\hspace{-0.1cm} F_{B}, a = d  \\
      C_{B}, &  \hspace{-0.2cm} \text{if}\ x=F_{B}, a \in \mathcal{A} \\
      0, & \hspace{-0.2cm} \text{if}\ x=F_T.
    \end{cases} \nonumber
  \end{equation}
The optimal policy $\pi^*_{B}$ is induced as a result of the minimization of $c^{\pi_{B}}(x,a)$.
The constraint on the SP's confidence in $s$ is enforced with a large instantaneous cost $C_{B}$ for reaching state $F_{B}$, which is ideally infinite. Assuming that the system follows the optimal policy, data release actions resulting in a transition to $F_{B}$ with a large-cost $C_{B}$ would not be selected by the minimization problem. See the proof of Theorem \ref{thm:convergence} in Appendix \ref{apx:ThmProof}.
The overall strategy for belief update is represented by the Bayes’ operator as follows:
\begin{equation}
    \Phi^{\pi_{B}}(x,z,a)  =    \begin{cases}
      \Phi^{\pi_{B}}(\beta,z,a),  &  \text{if}\ x=\beta \in \mathbb{P}(\mathcal{B}), a \in \mathcal{A} \\
      F_T, &  \text{if}\ x=\beta \in \mathbb{P}(\mathcal{B}), a = d \\
      F_T, &  \text{if}\ x=F_T.
    \end{cases} \nonumber
  \end{equation}
Since the user has access to all the information that the SP has, it can perfectly track his beliefs. Hence, the user decides her own policy facilitating the SP's detection strategy, episodic behavior and belief.

According to her strategy, the user checks whether the selected optimal action is the stopping action $d$. If so, she receives a cost determined by the current error probability of $u$ with penalty $\lambda$, then transitions to the terminal state and ends the episode. If not, she checks whether the SP's belief on any secret exceeds ${L_{B}}$. If the user is in the \textit{forbidden-state} she receives a large-cost $C_{B}$; otherwise, either she receives a time cost 1 or terminal state cost 0 depending on her state. 
If the terminal state has not already been reached and stopping action has not been taken at the moment, the user updates the SP's belief as in (\ref{BeliefUpdate}); otherwise she updates the state to the final state $x=F_T$. Using the condition (\ref{eqn:thm1}) in Theorem \ref{thm:convergence}, we write the Bellman equation induced by the optimal policy $\pi^*_{B}$ as \cite{Puterman},

\vspace{-0.3cm}
\begin{align}
    V(x) = \hspace{-0.3cm} \min \limits_{a\in \mathcal{A}\cup \{d\}}\{c^{\pi_{B}}(x,a)+\mathbb{E}[V(\Phi^{\pi_{B}}(x,z,a))]\}, \forall x \in \mathbb{P}(\mathcal{B}). \label{eqn:Bellman}
\end{align}

\vspace{-0.3cm}
The objective is to find a policy $\pi^*_{B}$ that optimizes the cost function. The proposed POMDP has a continuous state space due to belief state and continuous action probabilities. Finding optimal policies for continuous state and action is PSPACE-hard \cite{PSPACEhard}. In practice, to solve them by classical finite-state MDP methods, e.g., value iteration, policy iteration and gradient-based methods, belief discretization is required \cite{Tamas}. While a finer discretization gets closer to the optimal solution, it expands the state space; hence, the problem complexity. Hence, we use A2C-DRL to numerically solve the continuous state and action space MDP in Section \ref{sec:NumericalResults}.

In addition to the confidence-based privacy, we also consider a MI privacy policy in Section \ref{sec:MI}.

\vspace{-0.2cm}
\section{MI as Privacy Constraint}
\vspace{-0.1cm}
\label{sec:MI}
In this section, we consider a scenario, in which the user is interested in limiting the information leakage about the sensitive information in an average sense, rather than hiding its true value. For instance, the SP might be confused about the true secret; however, he might still have an idea about which secret values are unlikely. More concretely, consider a secret r.v. with alphabet size of three, e.g., $\mathcal{U}=\{1,2,3\}$. From the perspective of confidence, the belief of $\beta(U=1)=1/2$, $\beta(U=2)=1/4$, $\beta(U=3)=1/4$ would be the same as $\beta(U=1)=1/2$, $\beta(U=2)=1/2$, $\beta(U=3)=0$. While the latter clearly has additional information about the secret resulting in reduced uncertainty.
We tackle this issue by measuring the privacy by the MI between the secret variable $S$ and the observation history $\{Z^{t},A^{t}\}$ for $t \leq \tau$. According to her policy, the user wants to minimize the error on useful information as quickly as possible while keeping the total MI between the secret and the observations below a prescribed level, i.e., $\forall Z \in \mathcal{Z}$ and $\forall A \in \mathcal{A}$,

\vspace{-0.5cm}
\begin{align}
\centering
    &\minimize  \ \ \mathbb{E}[\tau]+\lambda P_{err}(u) \label{eqn:ProblemPP}\\
    &\subjectto \ \ \ \ \ \ \ \ \ I(S;Z^{t},A^{t}) < L_{MI}, \ \ \forall t \leq \tau, \forall S \in \mathcal{S} \nonumber
\end{align}
where $L_{MI}$ is a scalar of the user's choice. 

MI is commonly used both as a privacy and a utility measure in the literature \cite{BizBook, TIFS, Funnel}. Here, it is used as a privacy measure to control PUT between the useful variable and the secret. Due to the MI-based privacy constraint in (\ref{eqn:ProblemPP}), we call this policy \textit{MI-privacy data release policy} (MI-PDRP), $\pi_{MI}$. MI between $S$ and $(Z^T,A^T)$ over time $T$ is given by

\vspace{-0.4cm}
\begin{align}
     I(S;Z^T,A^T)& =\sum\limits_{t=1}^T I(S;Z_t,A_t|Z^{t-1},A^{t-1}).\label{eq:MIchain}
\end{align}
\begin{theorem}
The instantaneous MI cost between the secret and the observations induced by policy $\pi_{MI}$ at time $t$ can be written as:

\vspace{-0.4cm}
\begin{align}
    I^{\pi_{MI}}(&S;Z_t,A_t|\beta)
    =-\sum\limits_{s,u,z_t,a_t}q(z_t|a_t,s,u)\pi(a_t|\beta)\beta(s,u) \nonumber \\
    &\times \log \frac{\sum\limits_{\tilde{u}}q(z_t|a_t,s,\tilde{u})\pi(a_t|\beta)\beta(s,\tilde{u})}{\beta(s)\sum\limits_{\bar{s},\bar{u}}q(z_t|a_t,\bar{s},\bar{u})\pi(a_t|\beta)\beta(\bar{s},\bar{u})}. \label{eqn:instantaneousMI}
\end{align}
\label{thm:SeqDecisionMI}
\end{theorem}
See Appendix \ref{apx:SequentialDecision} for the proof.

As before, we define the state in three stages, i.e., the belief, the \textit{forbidden-MI-state} as $F_{MI}:=\{\beta_t(s): I^{\pi_{MI}}(S;Z^{t},A^{t}) \geq L_{MI}, \forall t \leq \tau \}$ for $F_{MI}\subseteq\mathbb{P}(\mathcal{B})$, where the constraint in (\ref{eqn:ProblemPP}) is violated, and the final state $F_T$ in which the episode terminates. As before, we will use $F_{MI}$ to denote both the forbidden state and the set of forbidden states for convenience.
We define an instantaneous cost function, $c^{\pi_{MI}}(x,a)$, for current state $x\in \mathcal{X}=\mathbb{P}(\mathcal{B}) \cup \{F_T\}$ and action $a \in \mathcal{A} \cup \{d\}$, which induces the optimal MI-PDRP $\pi^*_{MI}$ when minimized:
\begin{equation}
 \hspace{-0.1cm}   c^{\pi_{MI}}(x,a) \hspace{-0.1cm} = \hspace{-0.15cm}
    \begin{cases}
      1, \hspace{-0.2cm}  & \hspace{-0.3cm} \text{if}\ x\hspace{-0.05cm}=\hspace{-0.05cm}\beta \in \mathbb{P}(\mathcal{B}) \hspace{-0.1cm} \setminus \hspace{-0.1cm} F_{MI},  a \in \mathcal{A} \\
      \min \limits_{u \in \mathcal{U}}(1-\beta(u))\lambda, & \hspace{-0.3cm} \text{if}\ x\hspace{-0.05cm}=\hspace{-0.05cm}\beta \in \mathbb{P}(\mathcal{B}) \hspace{-0.1cm} \setminus \hspace{-0.1cm} F_{MI},   a = d  \\
      C_{MI}, &  \hspace{-0.3cm} \text{if}\ x\hspace{-0.05cm}=\hspace{-0.05cm}F_{MI}, a \in \mathcal{A} \\
      0, & \hspace{-0.3cm} \text{if}\ x\hspace{-0.05cm}=\hspace{-0.05cm}F_T.
    \end{cases} \nonumber
  \end{equation}
The constraint on the total MI leakage from $S$ is enforced with a large-cost $C_{MI}$ for state $F_{MI}$. Assuming that the system follows the optimal MI-PDRP $\pi^*_{MI}$, $F_{MI}$ would not be visited at all.
The overall strategy for belief update is represented by the Bayes’ operator as follows:
\begin{equation}
    \Phi^{\pi_{MI}}(x,z,a) \hspace{-0.1cm} = \hspace{-0.1cm}
    \begin{cases}
      \Phi^{\pi_{MI}}(\beta,z,a),  & \hspace{-0.3cm} \text{if}\ x=\beta \in \mathbb{P}(\mathcal{B}), a \in \mathcal{A}, \\
      F_T, & \hspace{-0.3cm} \text{if}\ x=\beta \in \mathbb{P}(\mathcal{B}), a = d, \\
      F_{MI}, & \hspace{-0.3cm} \text{if}\ x=\beta(s,u) \in \mathbb{P}(\mathcal{B}) \\
      & \hspace{-0.3cm} \sum \limits^{t}_{i=1}I^{\pi}(S;Z_i,A_i|\beta_i) \geq L_{MI},\\
      F_T, & \hspace{-0.3cm} \text{if}\ x=F_T.
    \end{cases} \nonumber
  \end{equation}

Theorem \ref{thm:convergence} holds for (\ref{eqn:ProblemPP}) when we replace $\{c^{\pi_{B}}, \Phi^{\pi_{B}}, F_B\}$ with $\{c^{\pi_{MI}}, \Phi^{\pi_{MI}}, F_{MI}\}$, and provides a lower bound for the value function $V^*$ for all $\beta \in \mathbb{P}(\mathcal{B})$. 
Hence, to find the policy $\pi_{MI}^*$, we solve the Bellman equation (\ref{eqn:Bellman}) using RL for $c^{\pi_{B}}$ and $ \Phi^{\pi_{B}}$. 
This policy minimizes the SP's error on the true value of $u$ in the quickest way while constraining the MI leakage from not only true secret $s$ but all possible values for $S$. 

\vspace{-0.4cm}
\subsection{Estimating MI}
Exact computation of MI is possible when the data distribution is known. However, in most practical scenarios, the user's data distribution is not known or it is inaccurate. Hence, we approximate $I(S;Z^{T},A^{T})$ via a variational representation which is inspired by Barber-Agakov MI estimation for single letter MI \cite{BA_MI}. 
Since (\ref{eq:MIchain}) is history-dependent, we modify this variational bound to a history dependent expression as follows:
\begin{align}
     I(&S;Z_t,A_t|Z^{t-1},A^{t-1}) && \nonumber \\
    & = H(S|Z^{t-1},A^{t-1}) && \hspace{-0.5cm} - H(S|Z^{t},A^{t})  \label{eqn:DefMI}\\
    & = H(S|Z^{t-1},A^{t-1}) && \hspace{-0.5cm} + \text{D}(P(S|Z^{t},A^{t})||Q(S|Z^{t},A^{t})) \nonumber \\ 
    & && \hspace{-0.5cm} + \mathbb{E}[\log Q(S|Z^{t},A^{t})] \label{eqn:identityMI} \\
    & = H(S|Z^{t-1},A^{t-1}) && \hspace{-0.5cm} + \max \limits_{Q(S|Z^{t},A^{t})}\mathbb{E}[\log Q(S|Z^{t},A^{t})] \label{eqn:BAMI}
\end{align}
where (\ref{eqn:DefMI}) follows from the definition of MI, (\ref{eqn:identityMI}) holds for any distribution $Q(S|Z^{t},A^{t})$ over $\mathcal{S}$ given the values in $\mathcal{Z}^t\times \mathcal{A}^t$, which represents what the belief would be after observing $(A_t,Z_t)$, and (\ref{eqn:BAMI}) follows from the fact that maximum is attained when $Q(S|Z^{t},A^{t})=P(S|Z^{t},A^{t})$.

Given $(Z^{t-1},A^{t-1}) = (z^{t-1},a^{t-1})$, we can rewrite the variational representation for the MI conditioned on the neural estimation of the current belief $\hat{\beta}(S)=Q(S|Z^{t-1},A^{t-1})$ as
\begin{align}
    \hspace{-0.3cm} I(S;Z_t,A_t|\hat{\beta}) \hspace{-0.1cm} = \hspace{-0.1cm}  H(\hat{\beta}(S)) + \hspace{-0.4cm} \max \limits_{Q(S|Z_{t},A_{t},\hat{\beta})} \hspace{-0.5cm} \mathbb{E}[\log Q(S|Z_{t},A_{t},\hat{\beta})], \label{eqn:BAMI2}
\end{align}
where $H(\hat{\beta}(S))=-\sum \limits_{s \in \mathcal{S}}\hat{\beta}(s)\log\hat{\beta}(s)$, and the expectation is with respect to $(S,Z_t,A_t) \sim \hat{\beta}(S),\pi(A_t|\hat{\beta}),{q}(Z_t|A_t,S,U)$. Since the current belief realization is known to both the user and the SP, $H(\hat{\beta}(S))$ is a constant. Numerical estimation of the MI via neural networks is explained in Section \ref{sec:NumericalResultsMI}.

\vspace{-0.3cm}
\section{Numerical Results}
\begin{figure}[pt]
\centering
\includegraphics[width=8.8cm]{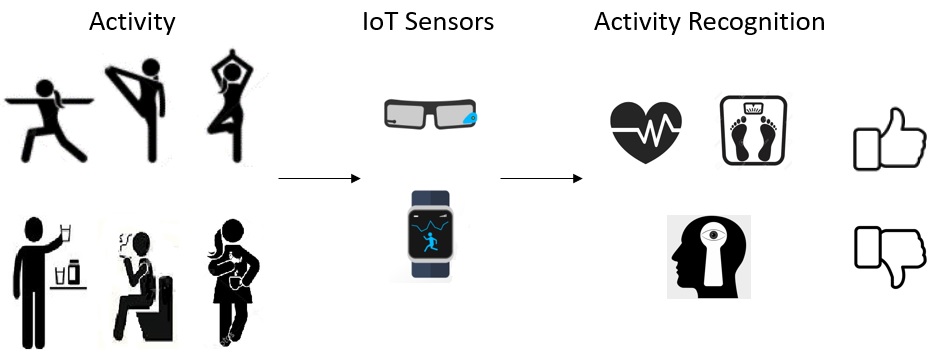}
\caption{Activity recognition with wearable IoT devices does not only infer physical exercise but also sensitive daily habits.} 
\label{fig:Activity}
\end{figure}

\begin{figure*}[ht]
\center
    \subfloat[]{\label{fig:SynB_T}\includegraphics[width=0.4\textwidth]{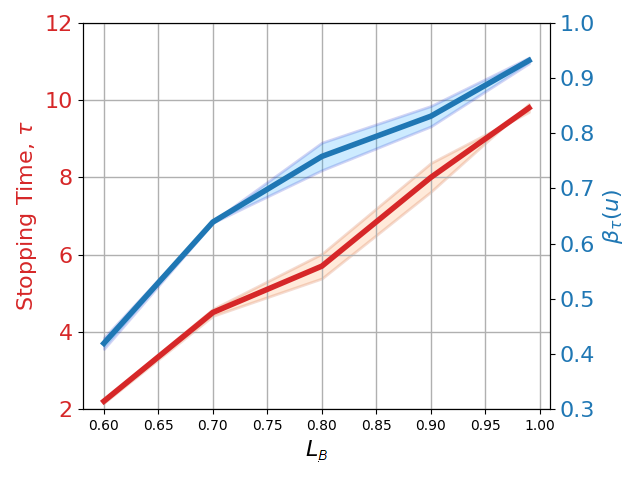}}\hspace{0.3cm}
    \subfloat[]{\label{fig:SynB_Acc}\includegraphics[width=0.44\textwidth]{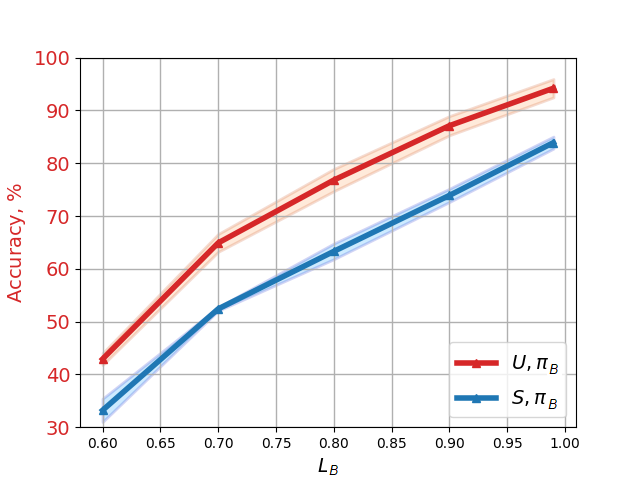}}
    \vfill 
    \subfloat[]{\label{fig:SynMI_T}\includegraphics[width=0.4\textwidth]{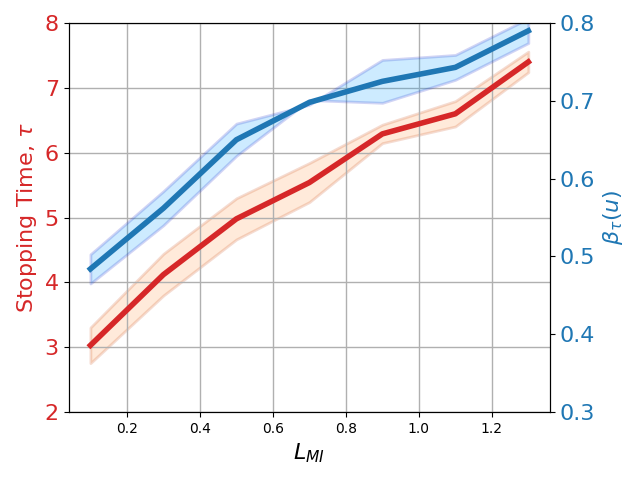}}\hspace{0.3cm}
    \subfloat[]{\label{fig:SynMI_Acc}\includegraphics[width=0.44\textwidth]{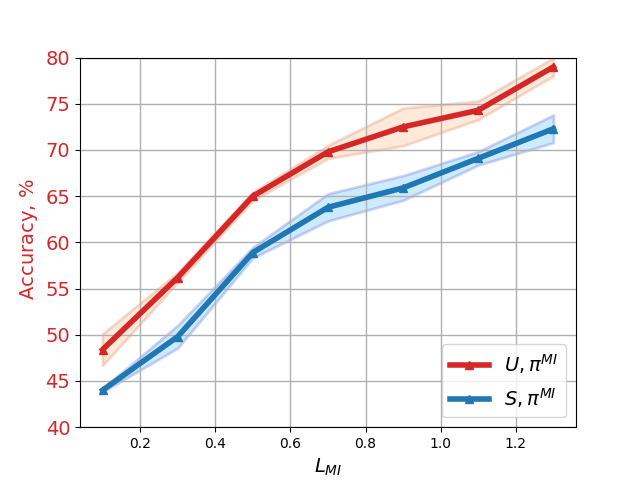}} 
     \vspace{-0.1cm}
    \caption{Belief-PDRP's, $\pi_{B}$, (a) stopping time $\tau$ and ${\beta}(u)$, and (b) SP's accuracy for the secret and the useful information with respect to $L_{B}$, and MI-PDRP's, $\pi_{MI}$, (c) stopping time $\tau$ and ${\beta}(u)$, and (d) SP's accuracy for the secret and the useful information with respect to $L_{MI}$.}
    \label{fig:SyntheticResults}
\end{figure*}

In this section, we present our results for both synthetic data and human activity privacy use-cases. In the former we assume that all the distributions of the DRM are known by both the user and the SP, while in the latter, these distributions are learnt from a real dataset. 
In human activity privacy use-case, we focus on the sensors in wearable devices as an example of DRMs, and their measurements as time-series data. In this scenario, the user shares sensor readings of her wearable device with the SP, while performing physical activities, with the goal of tracking the type and duration of her activities. However, as in Fig. \ref{fig:Activity}, not only useful activities, such as exercise type, but also sensitive activities, such as smoking, drinking or eating habits, can be inferred from these readings, which the user may not want to share with the SP as the SP can exploit such information for commercial benefit at the detriment of the user. Hence, the user shares a single sensor reading from among multiple sensors at a time such that the useful activity is revealed to the SP while his confidence in the sensitive activity is kept hidden at a pre-defined level.

The POMDP formulation in Section \ref{sec:POMDPFormulation} enable us to numerically approximate the proposed policies using RL. In RL, an agent discovers the best action to take in a particular state by receiving instant rewards or costs from the environment \cite{SuttonBarto}.
POMDPs with continuous belief and action spaces are difficult to solve numerically by using classical MDP solution methods. Actor-critic RL algorithms combine the advantages of value-based (critic-only) and policy-based (actor-only) methods, such as low variance and continuous action probability producing capability. Therefore, we use A2C-DRL for the numerical evaluation of our problem.

\subsection{A2C-DRL}
\label{sec:NumericalResultsA2C}
In the A2C-DRL algorithm, the actor represents the policy structure and the critic estimates the value function \cite{SuttonBarto}. In our setting, we parameterize the value function by the parameter vector $\theta \in \Theta$ as $V_{\theta}(x)$, and the stochastic policy by $\xi \in \Xi$ as $\pi_{\boldsymbol{\xi}}$. The error between the critic's estimate and the target differing by one-step in time is called temporal difference (TD) error \cite{SurveyACRL}. The TD error for the experience tuple $(x_t,\pi(a_t|x_t),z_t,x_{t+1},c_t(x_t,a_t))$ is estimated as
\begin{align}
    \delta_t=c_t(x_t)+\gamma V_{\theta_t}(x_{t+1})-V_{\theta_t}(x_t),
\end{align}
where $c_t(x_t)+\gamma V_{\theta_t}(x_{t+1})$ is called the TD target, and $\gamma$ is a discount factor chosen close to $1$ to approximate the Bellman equation for our episodic MDP. Instead of using the value functions in actor and critic updates, we use the advantage function to reduce the variance from the policy gradient. The advantage is approximated by TD error. Hence, the critic is updated by gradient ascent as:
\begin{align}
    \theta_{t+1}=\theta_t+\eta_t^c \nabla_{\theta}\ell_{c}(\theta_t),
\end{align}
where $\ell_c(\theta_t)=\delta_t^2$ is the critic loss, and $\eta_t^c$ is the learning rate of the critic at time $t$. The actor is updated similarly as,
\begin{align}
    \xi_{t+1}=\xi_t - \eta_t^a \nabla_{\xi}\ell_a(\xi_t),
\end{align}
where $\ell_a(\xi_t)=-\ln(\pi(a_t|x_t,\xi_t))\delta_t$ is the actor loss and $\eta_t^a$ is the actor's learning rate. 
In implementation, we represent the actor and critic by fully connected deep neural networks (DNNs) with two hidden layers of 256 nodes and \textit{Leaky-ReLU} activation.
The critic DNN takes the current state $x$ of size $N\times M$ as input, and outputs the corresponding state value for the current action probabilities $V_{\theta}^{\xi}(x)$. The actor takes the state as input, and outputs the corresponding action probabilities $\{\xi^0, \dots, \xi^{|\mathcal{\mathcal{A}}|}\}$ from a \textit{softmax} layer for $a \in \mathcal{A} \cup d$. 

\vspace{-0.3cm}
\subsection{Synthetic Data Use-Case}
The synthetic data scenario represents the situations where the probability distributions of DRMs and belief update rules are known by both the user and the SP, while only the actions are learned by the privacy mechanism.  

We create a dataset for $|\mathcal{A}\cup \{d\}|$=4, $|\mathcal{S}|$=3, $|\mathcal{U}|$=3,  $|\mathcal{Z}|$=50 and uniformly distributed $S$ and $U$, and $L_{B}\in\{0.6,0.7,0.8,0.9,0.99\}$. Observation probabilities are selected such that each action distinguishes a different pair of hypotheses well for both $S$ and $U$. For example, we created a matrix with each row representing the conditional distribution of $z$ for different $(a,s,u)$ realizations. For sensor $a$ = $0$, we used $\mathcal{N}(0,\sigma_j)$ for $(s,u)$ = $\{(0,0), (0,1), (0,2), (1,0), (1,1), (1,2)\}$, $\mathcal{N}(1,\sigma_j)$ for $(s,u)$ = $(2,0)$, $\mathcal{N}(2,\sigma_j)$ for $(s,u)$ = $(2,1)$, and $\mathcal{N}(3,\sigma_j)$ for $(s,u)$ = $(2,2)$, and we normalized through the columns representing $z$. Here, $\sigma_j$'s are chosen randomly from the interval $[0.5,1.5]$ for each $(a,s,u)$ with index $j$=$\{1,..,N$$\times$$M$$\times$$|\mathcal{A}|\}$. This sensor discloses $s$=$2$ case more than the other secrets. Moreover, $a$=$1$ and $a$=$2$ reveal more information for $s$=$1$ and $s$=$0$ cases, respectively.

Fig. \ref{fig:SynB_T} shows the average stopping time $\tau$ and the maximum belief on $u$, ${\beta}(u)$, with respect to $L_{B}$ for the belief-PDRP, $\pi_{B}$. As the constraint on ${\beta}(s)$ is relaxed, the stopping time increases as well as the maximum ${\beta}(u)$. In Fig. \ref{fig:SynB_Acc}, on the other hand, we present the prediction accuracy of the true-useful activity $u$ from the belief calculation.
Red lines in Fig. \ref{fig:SynB_Acc} represent accuracy on $u$, and blue lines show the accuracy on $s$. The gap between the accuracy shows the effectiveness of the proposed policy $\pi_{B}$ in minimizing the SP's error probability of $u$ in the quickest way while keeping his confidence in $s$ below the threshold for the synthetic data.

Fig. \ref{fig:SynMI_T} shows the average stopping time $\tau$ and the maximum confidence in $u$, $\hat{\beta}(u)$, with respect to $L_{MI}$ for the MI-PDRP, $\pi_{MI}$. As before, when the constraint on MI is relaxed, the stopping time increases as well as the maximum $\hat{\beta}(u)$. In Fig. \ref{fig:SynMI_Acc}, 
red lines represent accuracy on $u$, and blue lines show the accuracy on $s$. Although $\pi_{MI}$ shows similar results with $\pi_{B}$, $\pi_{B}$ is more effective in hiding the true realization of $S$. This is because MI-PDRP provides PUT by constraining the statistics of all the realizations of $S$ rather than only the true realization.

\vspace{-0.6cm}
\subsection{Human Activity Privacy Use-Case}
\vspace{-0.1cm}
In human activity privacy scenario, We use \textit{smoking activity dataset} \cite{smokingDataset} which contains more than 40 hours of sensor measurements for activities, such as smoking while walking, drinking while standing, sitting etc. We use measurements from four selected sensors of a smartwatch, i.e., $|\mathcal{A}\cup \{d\}|=5$. Table \ref{tab:SmokingSensorsandActivities} shows these sensors and sensitive-useful activity pairs from the dataset.
We learn the probability distributions together with the actions from the real-world measurements.
\label{sec:NumericalResults}

{
\begin{table}[pt]
\captionsetup{justification=centering,format=myformat,labelsep=newline}
    \centering
\caption{S\textsc{elected} A\textsc{ctivities} \textsc{and} S\textsc{martwatch} S\textsc{ensors} \textsc{from} {S\textsc{moking} A\textsc{ctivity} D\textsc{ataset}}.}
    \label{tab:SmokingSensorsandActivities}
\begin{tabular}{lc|lc}
\hline
\textbf{Sensors:}    & \textbf{A} & \textbf{Activities:}   & \textbf{(S,U)} \\ \hline
Accelerometer        & 0          & Sitting                & (0,0)            \\
Gyroscope            & 1          & Standing               & (0,1)            \\
Magnetometer         & 2          & Walking                & (0,2)            \\
Linear-accelerometer & 3          & Sitting while smoking  & (1,0)            \\
                     &            & Standing while smoking & (1,1)            \\
                     &            & Walking while smoking  & (1,2)            \\
                     &            & Sitting while drinking & (2,0)            \\
                     &            & Standing while drinking & (2,1)   \\
                     \hline
\end{tabular}
\end{table}}

{
\begin{figure}[ht]
    \center
    \subfloat{\includegraphics[width=0.45\textwidth]{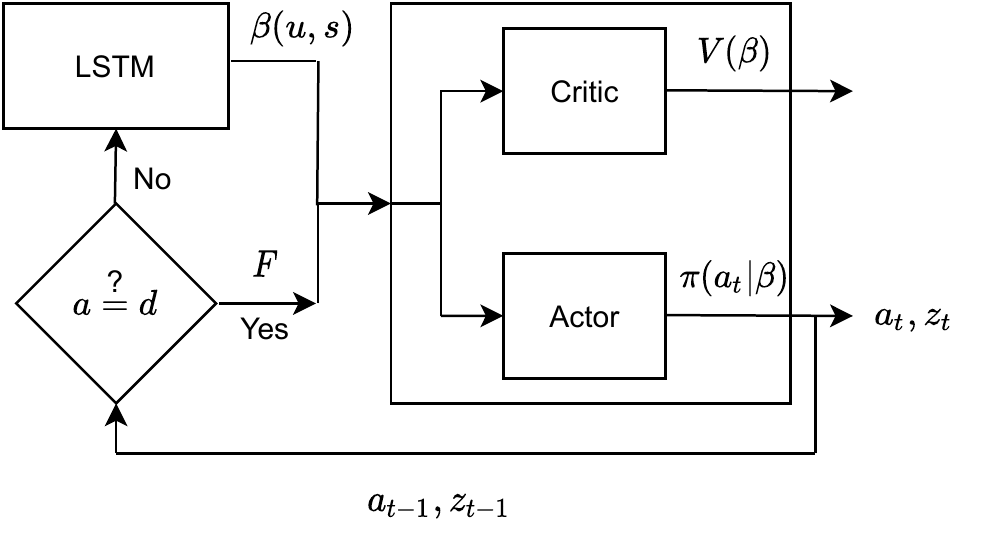}}
    \vspace{-0.1cm}
    \caption{A2C-DRL process for belief-PDRP, $\pi_{B}$.}\label{fig:BeliefNN}
\end{figure}}
{
\begin{figure*}[ht]
    \center
    \subfloat[]{\label{fig:Beliefa}\includegraphics[width=0.41\textwidth]{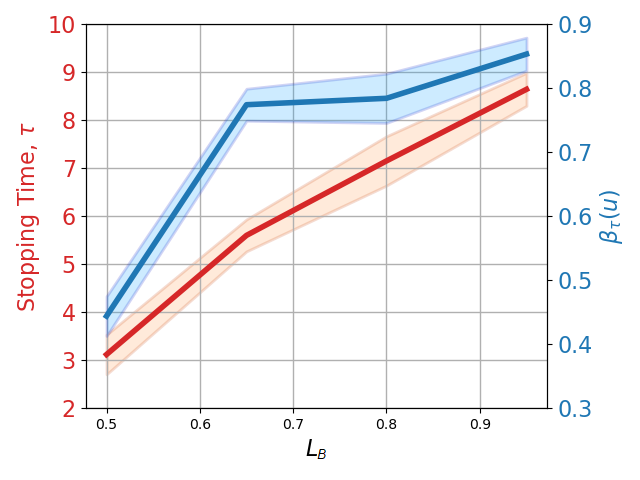}}
    \subfloat[]{\label{fig:Beliefb}\includegraphics[width=0.44\textwidth]{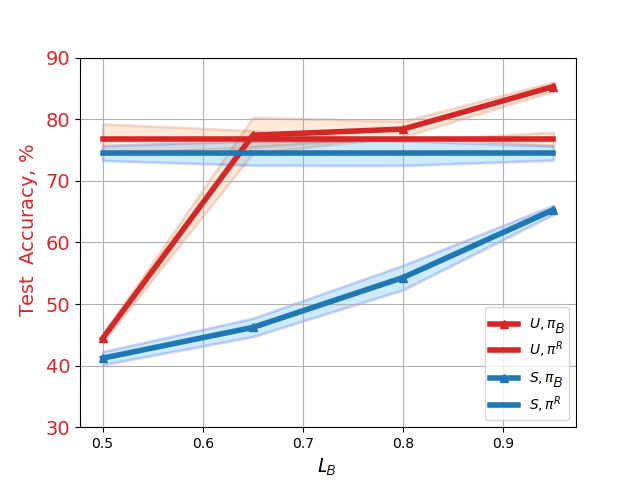}} 
    \vspace{-0.1cm}
    \caption{(a) Stopping time $\tau$ and $\hat{\beta}(u)$, and (b) SP's accuracy for the secret and the useful information with respect to $L_{B}$.}
    \label{fig:BeliefResults}
\end{figure*}}

\subsubsection{Numerical Results for Belief-PDRP, $\pi_{B}$}
\label{sec:NumericalResultsBelief}
In this section, we evaluate the PUT of the proposed optimal policy $\pi_{B}$ for smoking activity dataset. We model the SP by a long short-term memory (LSTM) recurrent neural network with parameters $\phi$, which predicts the true useful variable $u$ and secret $s$. The LSTM-based predictor has 2 layers with 128 nodes and 2 look-backs, and inputs the past observations $\{z^{t-1},a^{t-1}\}$. The output is a probability distribution representing the belief vector $\hat{\beta}_{\phi}(S,U)$ obtained by minimizing a cross-entropy loss between $\hat{\beta}_{\phi}(S,U)$ and true values of $\{S,U\}$. This is equivalent to maximizing the log-likelihood of $\hat{\beta}_{\phi}(S,U)$, i.e.,
\begin{align}
    H(\beta,\hat{\beta})=-\sum_{s,u}\beta(s,u)\log(\hat{\beta}(s,u))=-\mathbb{E}_{s,u}[\log(\hat{\beta}(s,u))]. \nonumber
\end{align}
To train the LSTM SP beforehand, we split the training data into 3 portions. One is for pre-training the LSTM SP, which will be used during A2C-DRL, one is for online A2C-DRL training, and the last portion is to train an SP, i.e., LSTM predictor, for testing the performance of PUT with A2C-DRL. Let $\pi_R$ be a random policy with uniform action probabilities. We create observation pairs $\{Z_t,A_t\}$ for LSTM training by randomly sampling actions $A_t$ from $\pi_R$, and obtaining time-series $Z_t$ from the corresponding portion of the dataset. We also used $C_T=0.5$ for the time cost, and $\lambda=50$.

Fig. \ref{fig:BeliefNN} shows A2C-DRL process in which LSTM is used as an online state predictor from the past observations. The user checks if the termination action, i.e., $a_{t-1}=d$, has been taken, then she accordingly terminates the process. Otherwise she predicts the current belief with the LSTM network, and selects an action $a_t$ via the actor. The actor-critic network updates its parameters with the state value $V(\beta)$ and action probability $\pi(a_t|\beta)$ accordingly. Sensor reading $z_t$ is observed as per the selected action, and the observation pair $z_t,a_t$ is shared with the SP.

Fig. \ref{fig:Beliefa} shows the average stopping time $\tau$ and the predicted maximum belief on $u$, $\hat{\beta}(u)$, with respect to $L_{B}$ for the belief-PDRP, $\pi_{B}$. As the constraint on $\hat{\beta}(s)$ is relaxed, the stopping time increases as well as the maximum $\hat{\beta}(u)$. In Fig. \ref{fig:Beliefb}, on the other hand, no-PUT and PUT cases are compared in terms of prediction accuracy of the test SP on true-useful activity $u$ and the secret $s$, where accuracy of the SP for the randomly generated $A_t$ and corresponding $Z_t$ represents no-PUT case, while its accuracy for the A2C-DRL generated actions $A_t$ and $Z_t$ represents the PUT case. 
Red lines in Fig.  represent accuracy on $u$, and blue lines show the accuracy on $s$. The flat lines show the no-PUT case which does not depend on $L_{B}$, and the curved lines represent the PUT case. 
While the gap between the accuracy of $u$ and $s$ is very low for random policy (no-PUT case), it is very large for $\pi_{B}$ (PUT case). This shows the effectiveness of the proposed policy $\pi_{B}$ in minimizing the SP's error probability of $u$ in the quickest way while keeping his confidence in $s$ below the threshold. On the other hand, generating random actions from a random policy does not yield a sophisticated strategy to reveal $u$ and hide $s$. The largest gap, i.e., the best performance of $\pi_{B}$, occurs at $L_{B}=0.65$ for $\pi_{B}$.
See Appendix \ref{apx:Breakdown} for a detailed breakdown of the accuracy in $U$ and $S$ realizations.

\begin{figure}[t]
    \center
    \subfloat{\includegraphics[width=0.45\textwidth]{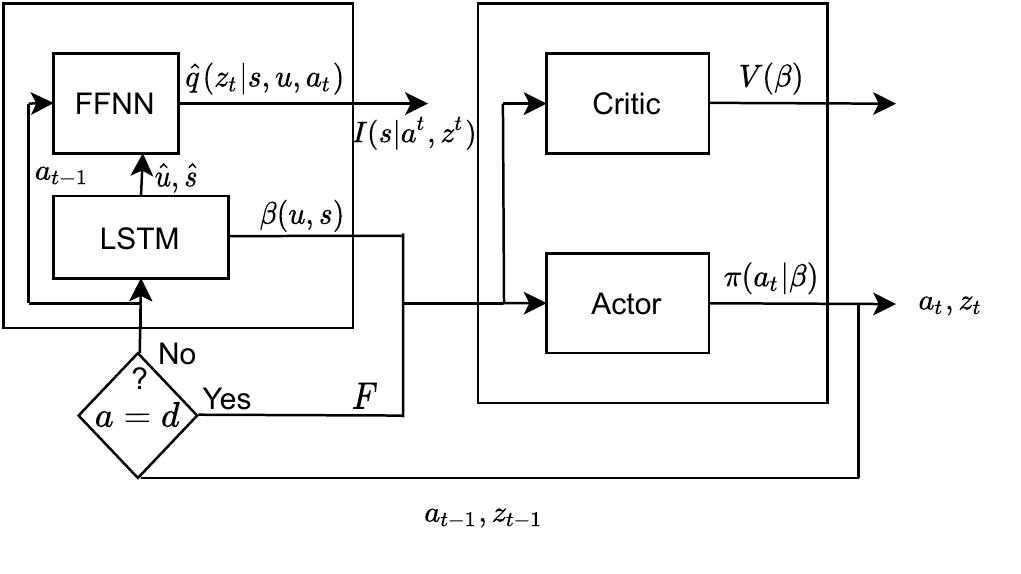}}
    \vspace{-0.1cm}
    \caption{A2C-DRL process for MI-PDRP, $\pi_{MI}$.}
    \label{fig:MINN}
  \end{figure}


\begin{figure*}[ht]
    \center
    \subfloat[]{\label{fig:MIa}\includegraphics[width=0.41\textwidth]{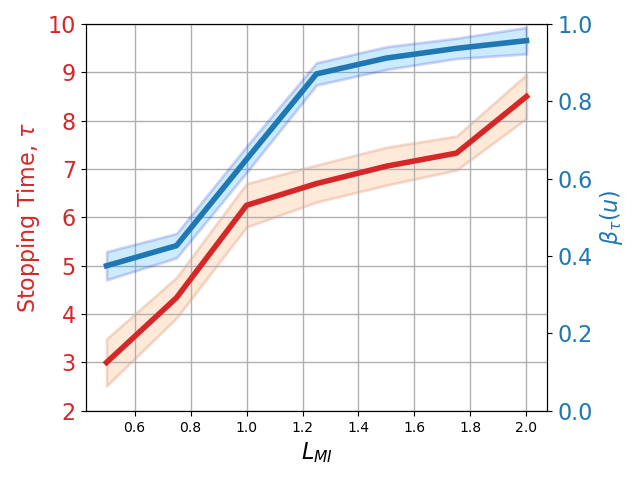}}
    \subfloat[]{\label{fig:MIb}\includegraphics[width=0.44\textwidth]{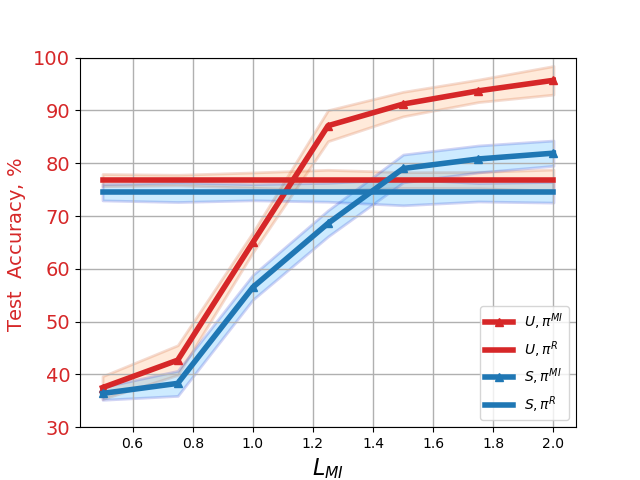}}
    \vspace{-0.1cm}
    \caption{(a) Stopping time $\tau$ and $\hat{\beta}(u)$ and (c) SP's accuracy for the secret and the useful information with respect to $L_{MI}$.}
    \label{fig:MIResults}
  \end{figure*}

\subsubsection{Numerical Results for MI-PDRP, $\pi_{MI}$}
\label{sec:NumericalResultsMI}
In this section, we model the SP using two components; one is an LSTM-based belief predictor with 2 layers of 128 nodes and 2 look-backs, and the other one is a feed-forward neural network (FFNN)-based observation generator with 3 layers of 256 nodes, where the output determines the mean $\mu$ and standard deviation $\sigma$ of a Gaussian distribution. As before, we use $C_T=0.5$ for the time cost, and $\lambda=50$.

As in Section \ref{sec:NumericalResultsBelief}, we train the LSTM network with parameters $\phi$ by minimizing a cross-entropy loss between the observations $\{Z^{t-1},A^{t-1}\}$ and $\{S,U\}$, which is equivalent to maximizing the log-likelihood of $\hat{\beta}_{\phi}(S,U)$. As a result, KL divergence between the real belief distribution $\beta$ and the predicted distribution $\hat{\beta}_{\phi}$ goes to zero when the log-likelihood is maximized \cite{BA_MI}. In addition, we estimate ${q}(Z_t|A_t,S,U)$, which is represented by a Gaussian distribution,

\vspace{-0.5cm}
\begin{equation}
    \hat{q}(Z_t|A_t,S,U) = \mathcal{N}(Z_t|(\mu,\Sigma))=f_{\psi}(A_t,S,U),
\end{equation}
where $(\mu,\sigma)$ are determined by a FFNN ${f}_{\psi}$ by maximizing its log-likelihood. During A2C-DRL, we sample observations $Z_t$ and $A_t$ to calculate the variational bound for MI using the pre-trained FFNN and LSTM networks which satisfy the maximization in (\ref{eqn:BAMI2}). We approximate the MI by sampling $k$ observations $\{z_t^{i},a_t^{i}\}_i^k \sim \hat{q}(z_t|a_t,\hat{s},\hat{u}),\pi_{MI}(a_t|\hat{\beta})$, and using the predictions for the next $k$ belief states $\{Q(s|z^i_t,a^i_t,\hat{\beta})\}_i^k$ as follows:

\vspace{-0.5cm}
\begin{align}
    \hat{I}(&S;Z_t,A_t|\phi,\psi) \nonumber\\ 
    & = H(\hat{\beta}_{\phi}) + \frac{1}{n}\sum \limits_{j=1}^{n} \Big[ \frac{1}{k}\sum \limits_{i=1}^{k} \log[Q_{\psi}((\hat{s}^j|z^i_t,a^i_t,\hat{\beta}_{\phi}))] \Big], \label{eqn:EstMI}
\end{align}

\vspace{-0.3cm}\hspace{-0.45cm}
 where $\hat{s}^j$ is a realization of $s$ sampled from the predicted belief vector $\hat{\beta}_{\phi}(s)$.
Fig. \ref{fig:MINN} illustrates A2C-DRL process with belief and MI calculation using pre-trained LSTM and FFNN. The user checks if the termination action, i.e., $a_{t-1}=d$, has been taken. If so, she accordingly terminates the process. Otherwise, she predicts the current belief from the previous observations using the LSTM network, and takes action $a_t$. The actor-critic network updates its parameters with the state value $V(\beta)$ and action probability $\pi(a_t|\beta)$ accordingly. Sensor measurement is observed as per the selected action, and the observation pair $z_t,a_t$ is shared with the SP. $\hat{I}(\hat{S}|A^t,Z^t|\beta_t)$ is calculated by the SP using previous action $a_{t-1}$ and $(\hat{s},\hat{u})$ according to (\ref{eqn:EstMI}).

Fig. \ref{fig:MIa} shows the average stopping time $\tau$ and the maximum confidence in $u$, $\hat{\beta}(u)$, with respect to $L_{MI}$ for the MI-PDRP, $\pi_{MI}$. As the constraint on MI is relaxed, the stopping time increases as well as the maximum $\hat{\beta}(u)$. In Fig. \ref{fig:MIb}, activity prediction accuracy of the test SP for observations $(Z_t,A_t)$ generated by random policy $\pi_R$ and $\pi_{MI}$ are compared.
Red lines in Fig. \ref{fig:MIResults} represent accuracy on $u$, and blue lines show the accuracy on $s$. Similarly to Section \ref{sec:NumericalResultsBelief}, the gap between the accuracy of $u$ and $s$ is very low for random policy, while it is large for $\pi_{MI}$. This shows that the proposed policy $\pi_{MI}$ minimizes the SP's error probability of $u$ in a speedy manner while keeping the information leakage from $s$ below the threshold. Although $\pi_{MI}$ shows similar results with $\pi_{B}$, $\pi_{B}$ is more effective in hiding the true realization of $S$. This is because MI-PDRP provides PUT by constraining the statistics of all the realizations of $S$ rather than only the true realization.
The largest gap in Fig \ref{fig:MIResults}, i.e., the best performance of $\pi_{MI}$, occurs at $L_{MI}=1.2$ for $\pi_{MI}$.
See Appendix \ref{apx:Breakdown} for a detailed breakdown of the accuracy in $U$ and $S$ realizations.

\vspace{-0.5cm}
\section{Conclusion}
\vspace{-0.2cm}
\label{sec:Conclusion}
We studied the PUT in time-series data release to an SP. The goal of the user is to reveal the true value of a latent utility variable, while keeping the secret variable private from the SP. In a sense, the SP is the legitimate receiver for the utility variable, while acting as the adversary for the sensitive variable. 
In particular, we measured the utility by
the confidence of the SP in the latent useful information.
For privacy, we considered both the confidence
of the SP on the sensitive information and the MI between the sensitive variable and the revealed measurements.
We proposed active sequential data release policies to minimize the error probability on the true useful variable in a speedy manner, while constraining the confidence of the SP or the MI leakage for the secret variable. We provided a POMDP formulation
of the problem, and used A2C-DRL for numerical
evaluations. 
Utilizing  DNNs,  we numerically evaluated the PUT curve of the proposed policies for \textit{smoking activity dataset}, where useful and sensitive activities are revealed to the adversary through smartwatch sensors selected by the user. We examined the effectiveness of the optimal belief-PDRP and MI-PDRP using an LSTM-based adversary network. According to the numerical results, we have seen that the proposed data release policies provide significant privacy advantage compared to random sensor selection. We have also seen that constraining the MI does not necessarily hide the true value of the secret at the same level as the belief-PDRP. However, this
approach may be more useful when the objective is not necessarily
to hide the true value of the secret, but limit the knowledge of the SP in an average sense. We have also shown that decision time gets longer when the constraint on the secret is relaxed.

\vspace{-0.2cm}
\appendices
\begin{table*}[ht]
\centering
\captionsetup{justification=centering,format=myformat,labelsep=newline}
\caption{A\textsc{dversary} A\textsc{ccuracy} \textsc{for} \textsc{all} A\textsc{ctivities} U\textsc{nder} B\textsc{elief-privacy} \textsc{and} MI\textsc{-privacy} P\textsc{olicies}.}
    \label{tab:Breakdown}
\begin{tabular}{cccclll}
\multicolumn{1}{l}{\textbf{Policy}} & \multicolumn{1}{c}{\textbf{Constraint}} & \multicolumn{1}{c}{\textbf{$\tau$/$\hat{\beta}(U)$}} & \multicolumn{1}{l}{\textbf{Acc. U}} & \textbf{Acc. U=0 / U=1 / U=2} & \textbf{Acc. S} & \textbf{Acc. S=0 / S=1 / S=2} \\ 
\hline
\multirow{4}{*}{$\pi_{B}$}   &0.5  & 3.12 / \%43               &  \%44.4      & \%60.67 / \%53.44 / \%5.65   & \%41.2 & \%24.58 / \%23.6 / \%4.58         \\
                                 &0.65 &  5.6  / \%74              &  \%77.4      & \%78.9 / \%69.15 / \%88.21   & \%46.2 & \%55.32 / \%48.05 / \%5.15         \\
                                 &0.8  &  7.15 / \%77              &  \%78.4      & \%85.07 / \%61.99 / \%92.52  & \%54.3 &  \%68.77 / \%48.95 / \%34.5        \\
                                 &0.95 &  8.64 / \%81              &  \%85.3      & \%91.58 / \%71.23 / \%96.3   & \%65.3 & \%79.48 / \%73.07 / \%42.8         \\ \hline
\multirow{7}{*}{$\pi_{MI}$} &0.5  & 3 / \%38    &\%37.5  &\%47.01 / \%51.1 / \%0            &\%36.4  &\%46.53 / \%52.01 / \%0.74    \\
                            &0.75 & 4.34 / \%40      &\%42.7 &\%63.29 / \%49.17 / \%0           &\%38.3  &\%47.26 / \%53.74 / \%1.2      \\
                            &1    & 6.25 / \%62      &\%65   &\%87.26 / \%55.35 / \%46.7        &\%56.5  &\%47.48 / \%57.42 / \%13.83      \\
                            &1.25 & 6.7 / \%88       &\%87.1 &\%94 / \%80.8 / \%80.86           &\%68.6  &\%89.55 / \%65.63 / \%28.74      \\
                            &1.5  & 7.06 / \%91      &\%91.2 &\%97.62 / \%84 / \%92.16          &\%79    &\%93.02 / \%72.22 / \%34.67      \\
                            &1.75 & 7.8 / \%93      &\%93.7 &\%98.42 / \%88.8 / \%96.25        &\%80.8  &\%96.9 / \%96.38 / \%87.3      \\
                            &2    & 8.5 / \%94       &\%95.7 &\%99.45 / \%96.19 / \%97.78       &\%81.9  &\%98,3 / \%98.12 / \%92.64      \\ \hline
\end{tabular}
\end{table*}

\vspace{-0.3cm}
\section{Proof of Theorem \ref{thm:convergence}}
\label{apx:ThmProof}
\vspace{-0.1cm}
We assume that after the user takes the stopping action for data release, the system goes to a recursive termination state, denoted by $F_T$, and remains there forever. Hence, the state space is $\mathcal{X}=\mathbb{P}(\mathcal{B}) \cup \{F_T\}$. Let instantaneous cost of taking action $a \in \mathcal{A} \cup \{d\}$

\vspace{-0.4cm}
\begin{equation}
 \hspace{-0.1cm}   c^{\pi_{B}}(x,a) \hspace{-0.1cm} = \hspace{-0.1cm}
    \begin{cases}
      1, \hspace{-0.3cm}  & \hspace{-0.2cm} \text{if}\ x=\beta \in \mathbb{P}(\mathcal{B})\hspace{-0.1cm} \setminus \hspace{-0.1cm} F_{B}, a \in \mathcal{A} \\
      \min \limits_{u \in \mathcal{U}}(1-\beta(u))\lambda, & \hspace{-0.2cm} \text{if}\ x=\beta \in \mathbb{P}(\mathcal{B})\hspace{-0.1cm} \setminus\hspace{-0.1cm} F_{B}, a = d  \\
      C_{B}, &  \hspace{-0.2cm} \text{if}\ x=F_{B}, a \in \mathcal{A} \\
      0, & \hspace{-0.2cm} \text{if}\ x=F_T.
    \end{cases} \nonumber
  \end{equation}
  
  \vspace{-0.2cm}
The constraint on the adversary's confidence in $s$ is enforced with an instantaneous cost $C_{B}$ for state $F_{B}$, which is ideally infinity but can be applied as a very large scalar in practice. Assuming that the system follows the optimal policy, transition to $F_{B}$ with a very large cost $C_{B}$ would not be chosen by the minimization problem.
The overall strategy for belief update is represented by the Bayes’ operator as follows:

\vspace{-0.4cm}
\begin{equation}
    \Phi^{\pi_{B}}(x,z,a)  =    \begin{cases}
      \Phi^{\pi_{B}}(\beta,z,a),  &  \text{if}\ x=\beta \in \mathbb{P}(\mathcal{B}), a \in \mathcal{A} \\
      F_T, &  \text{if}\ x=\beta \in \mathbb{P}(\mathcal{B}), a = d \\
      F_T, &  \text{if}\ x=F_T.
    \end{cases} \nonumber
  \end{equation}
Using the instantaneous cost and state update, the condition $V(\beta) \leq \min\{C_T+\min\limits_{a \in \mathcal{A}}(\mathbb{T}^a V^*)(\beta), \min\limits_{u \in \mathcal{U}}\lambda C_T(1-\beta(u))\}$ is rewritten as

\vspace{-0.5cm}
\begin{align}
    &V(F_T)=0, \nonumber\\
    &V(x)\leq \min \limits_{a\in \mathcal{A}\cup \{d\}}\{T_c c(x,a)+\mathbb{E}[V(\Phi(x,z,a))]\}, \forall x \in \mathbb{P}(\mathcal{B}), \nonumber
\end{align}
as well as the state sequence at $t=0,1,2, \dots$ is denoted by
\vspace{-0.2cm}
\begin{align}
    &X_0=x, \nonumber\\
    &X_n=\Phi(X_{n-1},Z,A(n)), \forall n, n > 0.  \nonumber
\end{align}

\vspace{-0.2cm}
When the condition is written in terms of the state sequence of duration $N$ for the optimal policy $\pi^*$, we obtain

\vspace{-0.4cm}
\begin{equation}
    V(x) \leq C_T \mathbb{E}_{\pi^*}[\sum \limits_{n=0}^{N-1}c(X_n,A_n)] + \mathbb{E}_{\pi^*}[V(X_N)].
\end{equation}
Taking the limit as $N \rightarrow \infty$, we get

\vspace{-0.5cm}
\begin{align}
    V&(x) \nonumber\\
    & \leq C_T \mathbb{E}_{\pi^*}[\sum \limits_{n=0}^{\infty}c(X_n,A_n)] + \lim \limits_{N \rightarrow \infty}\mathbb{E}_{\pi^*}[V(X_N)] \label{a}\\
    & = C_T V^*(x) + \lim \limits_{N \rightarrow \infty} \mathbb{E}_{\pi^*}[X_N] \label{b}\\
    & = C_T V^*(x) + \hspace{-0.2cm} \lim \limits_{N \rightarrow \infty} \mathbb{E}_{\pi^*} \Big[ V(F_T)\boldsymbol{1}_{\{X_N=F_T\}}   \\  
    & + V(F_{B})\boldsymbol{1}_{\{X_N = F_{B}\}} + 
     V(X_N)\boldsymbol{1}_{\{X_N\neq F_T, X_N \neq F_{B}\}} \Big] \nonumber \\
    & = C_T V^*(x) + \lim \limits_{N \rightarrow \infty} \mathbb{E}_{\pi^*} \Big[ V(X_N)\boldsymbol{1}_{\{X_N\neq F_T\}}  \\ 
    & +  V(F_{B})\boldsymbol{1}_{\{X_N = F_{B}\}} \Big] \nonumber \\
    & \leq C_T V^*(x) + \hspace{-0.05cm} \lambda \hspace{-0.05cm} \lim \limits_{N \rightarrow \infty} \hspace{-0.05cm} \Big( \mathbb{P}_{\pi^*}[X_N \neq F_T] + \mathbb{P}_{\pi^*}[X_N = F_{B}] \Big) \label{c}\\
    & = C_T V^*(x), \label{d}
    \end{align}
where (\ref{a}) holds due to the monotone convergence theorem; (\ref{b}) follows from the definition of $V^*$; (\ref{c}) is due to the fact that for any $\beta \in \mathbb{P}(\mathcal{B})$, $V(\beta) \leq \min \limits_{u \in \mathcal{U}}\lambda (1-\beta(u)) \leq \lambda$; and (\ref{d}) holds since $\lambda \geq V^*(x) \geq \mathbb{E}_{\pi^*}[\tau]=\sum_{n=0}^{\infty}P_{\pi^*}(\tau > n) = \sum_{n=0}^{\infty}P_{\pi^*}(X_n \neq F_T)$, and the probability of the system following the optimal policy $\pi^*$ to transition to highest-cost state $F_{B}$ at $N$ is zero, i.e, $\lim \limits_{N \rightarrow \infty} \mathbb{P}_{\pi^*}[X_N = F_{B}]=0$.

\vspace{-0.4cm}
\section{Proof of Theorem \ref{thm:SeqDecisionMI}}
\label{apx:SequentialDecision}

\vspace{-0.2cm}
Consider a POMDP with the state space $\mathcal{X}=\mathbb{P}(\mathcal{B}) \cup \{F_T\}$. At time $t$, a decision maker observes $Z^{t-1},A^{t-1}$ and chooses an action $A_t \in \mathcal{A}\cup\{d\}$ as follows:
\begin{align}
    A_t=f_t(Z^{t-1},A^{t-1}),
\end{align}
where $\boldsymbol{f}=(f_1,f_2, \dots)$ is called the policy. Based on the conditional probability $\pi(A_t|A^{t-1},Z^{t-1})$ of taking this action, $Z_t \in \mathcal{Z}$ is observed and revealed by the sensor distribution $q(Z_t|A_t,S,U)$, and the state evolves to the next belief state. At each step, the system incurs a per-step cost 

\begin{align}
    &c(s,u,z^t,a^t;\boldsymbol{f}):=\nonumber \\
    &\log\frac{P^{\boldsymbol{f}}(Z_t=z_t,A_t=a_t|S=s,Z^{t-1} \hspace{-0.1cm} =z^{t-1},A^{t-1}\hspace{-0.1cm} =a^{t-1})}{P^{\boldsymbol{f}}(Z_t=z_t,A_t=a_t|Z^{t-1}=z^{t-1},A^{t-1}=a^{t-1})}.\nonumber
\end{align}
The objective is to find a policy $\boldsymbol{f}=(f_1,\dots,f_T)$ that minimizes the total cost given by $\frac{1}{T}\mathbb{E}^{\boldsymbol{f}}\Big [ \sum\limits_{t=1}^{T}c(S,U,Z^t,A^t;\boldsymbol{f}) \Big ]$,
where the expectation is taken with respect to the distributions induced by the policy $\boldsymbol{f}$.

Let $\boldsymbol{f}=(f_1,\dots,f_T)$ be $f_t(z^{t-1},a^{t-1})=\pi(\cdot|z^{t-1},a^{t-1})$. Then the following holds: 
\begin{align}
   &I^{\pi_{MI}}(S;Z_t,A_t|Z^{t-1},A^{t-1})  = \sum \limits_{s,u,z^t,a^t}P^{\pi_{MI}}(S,U,Z^t,A^t) \nonumber \\ 
   & \times \log \frac{P^{\pi_{MI}}(Z_t\hspace{-0.1cm}=\hspace{-0.1cm}z_t,A_t\hspace{-0.1cm}=\hspace{-0.1cm}a_t|S\hspace{-0.1cm}=\hspace{-0.1cm}s,Z^{t-1}\hspace{-0.1cm}=z^{t-1},A^{t-1} \hspace{-0.1cm} = a^{t-1})}{P^{\pi_{MI}}(Z_t=z_t,A_t=a_t|Z^{t-1}=z^{t-1},A^{t-1}=a^{t-1})} \nonumber \\
   & = \mathbb{E}^{\boldsymbol{f}} \Bigg [ \sum\limits_{t=1}^{T}c(S,U,Z^t,A^t;\boldsymbol{f}) \Bigg ]
\end{align}

The probability distribution on $(S,U,Z^T,A^T)$ induced by the decision policy $\boldsymbol{f}$ is given by
\begin{align}
    P^{\boldsymbol{f}}(S=s, &U=u, Z^T=z^T, A^T=a^T) \nonumber    \\
    &=P(s,u)q(z_1|a_1,s,u)\pi(a_1) \nonumber \\ &\times \prod\limits_{t=2}^T \Big[ q(z_t|a_t,s,u)\pi(a_t|z^{t-1},a^{t-1})\Big],
\end{align}
where $\pi(\cdot|z^{t-1},a^{t-1}) = f(z^{t-1},a^{t-1})$. Under the transformations described above, $P^{\boldsymbol{f}}$ and $P^{\pi_{MI}}$ are identical probability distributions. As a result, $\mathbb{E}^{\boldsymbol{f}} \Bigg [ \sum\limits_{t=1}^{T}c(S,U,Z^t,A^t;\boldsymbol{f}) \Bigg ]=I^{\pi_{MI}}(S;Z_t,A_t|Z^{t-1},A^{t-1})$. Hence, Theorem 3 holds.

\vspace{-0.4cm}
\section{Additional Simulation Results}
\label{apx:Breakdown}
In Table \ref{tab:Breakdown}, there is detailed performance of $\pi_{B}$ and $\pi_{MI}$ policies, where "Acc." represents accuracy. Individual accuracies for $U$ and $S$ show that all activities are revealed as the constraint is relaxed. On the other hand, $U=2$ and $S=2$ are almost completely hidden for low constraint level, but they are revealed faster then the other hypotheses. Moreover, $\pi_{B}$ and $\pi_{MI}$ policies reveal or hide different activities better due to the different characteristics of the activities. We also see the same results that Fig. \ref{fig:BeliefResults} and Fig. \ref{fig:MIResults} show, i.e., $\pi_{B}$ outperforms $\pi_{MI}$ in minimizing the error probability of $U$ in a speedy manner while keeping the secret below the pre-defined level.

\vspace{-0.5cm}
\bibliographystyle{IEEEtran}
\bibliography{IEEEexample}

\end{document}